\documentclass[apjl]{emulateapj}
\usepackage{amssymb,graphicx}
\usepackage{epsfig}
\usepackage{psfrag}
\usepackage[usenames]{color}
\usepackage{multirow}
\usepackage{rotating}
\usepackage{hyperref}
\usepackage{ragged2e}

\newcommand{\eqref}[1]{(\ref{#1})}

\shorttitle{Dynamical Capture Mergers of BHs with spinning NSs}
\shortauthors{East, Paschalidis \& Pretorius}

\begin{document}

\title{Eccentric mergers of black holes with spinning neutron stars}

\author{William E.\ East${}^1$, Vasileios Paschalidis${}^2$, and Frans Pretorius${}^2$}
\affil{
${}^1$Kavli Institute for Particle Astrophysics and Cosmology, Stanford University, SLAC National Accelerator Laboratory, Menlo Park, CA 94025, USA\\ 
${}^2$Department of Physics, Princeton University, Princeton, NJ 08544, USA}

\begin{abstract}

We study dynamical capture binary black hole--neutron star (BH--NS) mergers focusing on
the effects of the neutron star spin.  These events may arise in dense stellar
regions, such as globular clusters, where the majority of neutron stars are
expected to be rapidly rotating.  We initialize the BH--NS
systems with positions and velocities corresponding to marginally unbound
Newtonian orbits, and evolve them using general-relativistic hydrodynamical
simulations.  We find that even moderate spins can significantly increase the
amount of mass in unbound material. In some of the more extreme cases, there can
be up to a third of a solar mass in unbound matter. Similarly, large amounts of
tidally stripped material can remain bound and eventually accrete onto the 
BH---as much as a tenth of a solar mass in some cases. These simulations
demonstrate that it is important to treat neutron star spin in order to make
reliable predictions of the gravitational wave and electromagnetic transient
signals accompanying these sources.

\end{abstract}


\maketitle

\section{Introduction}

Next generation ground-based gravitational wave (GW) detectors such as
aLIGO~\citep{LIGO} 
are expected to reach
design sensitivity within the next few years. Among their most
promising sources are mergers of compact objects (COs), including black
hole--neutron star (BH--NS) binaries. BH--NS mergers are also proposed
short-hard gamma-ray burst (sGRB) engines
(e.g.~\citet{Meszaros:2006rc}), and may power other electromagnetic
(EM) transients either
preceding~\citep{Hansen:2000am,McWilliams:2011zi,Paschalidis:2013jsa}
or following~\citep{2012ApJ...746...48M} the merger.  These EM
counterparts to GWs could be observed by current and future
wide-field telescopes such as PTF~\citep{2009PASP..121.1334R} and
LSST~\citep{2012arXiv1211.0310L}.

Extracting maximum information from such ``multimessenger''
observations requires careful modeling of BH--NS mergers.  
Several studies of quasicircular BH--NS inspirals using numerical
relativity simulations have been performed, see,
e.g.,~\citet{cabllmn10,Etienne:2012te, Kyutoku:2013wxa,
Tanaka:2013ixa,Foucart:2014nda}.  While
quasicircular binaries may dominate the global rates of BH--NS
encounters in the Universe, recent
calculations~\citep{Kocsis:2011jy,lee2010,Samsing:2013kua} suggest
that in dense stellar regions, such as galactic nuclei and globular
clusters (GCs), CO binaries can form through dynamical capture and
merge with non-negligible eccentricities. 
Compared to quasi-circular inspirals, these emit more GW energy in the high 
luminosity, strong-field regime of general relativity, and small
changes in the energy of the binary at each pericenter passage can lead
to relatively large changes in the time between GW bursts---the
leading order GW observable. Hence these systems could be excellent
laboratories to test gravity and measure the internal structure of a NS 
(insofar as this affects the energy of the orbit, e.g. tidal excitation of 
f-modes). 
Rates of these events are highly uncertain, but have been estimated to be up to
$\sim100\ \rm{yr}^{-1}\ Gpc^{-3}$. To realize this rich potential
to learn about the Universe from eccentric mergers, it is irrelevant 
what their rates are compared to quasi-circular inspiral, only that
eccentric mergers occur frequently enough that some events could be
observable by aLIGO within its lifetime. However, new detection pipelines would 
be needed that are better adapted to the repeated-burst nature of eccentric GW
mergers for aLIGO to efficiently detect them~\citep{Tai:2014bfa}. For more
discussion of rates, distinguishing features, and detection issues for eccentric
encounters, see~\citet{2013PhRvD..87d3004E} and references therein.

Motivated by the above, \citet{ebhns_letter,bhns_astro_paper}
performed fully general-relativistic hydrodynamical (GR-HD)
simulations of dynamical-capture BH--NS mergers. These studies explored
the effects of impact parameter, BH spin, and NS equation of state (EOS)
on GW emission and post-merger BH disk and ejecta masses.
Here, we expand upon this work by including the effects of NS spin. 
To date the only simulations including spinning NSs focused on quasicircular NS--NS mergers,
e.g.~\citet{Tichy:2011gw,Bernuzzi:2013rza}, demonstrating that even
moderate spins can affect the dynamics.

NS spin has two main effects: (1) it modifies the star's structure,
making it less gravitationally bound; (2) it changes the orbital
dynamics, e.g., by shifting the effective innermost stable orbit
(ISO).  This can impact not only the GWs from CO mergers,
but also the amount of matter forming the BH accretion disk that
putatively powers a sGRB, and the amount of unbound 
matter that powers other EM transients, such as kilonovae.  There
may also be effects on pre-merger EM signals since the NS spin
determines the light-cylinder radius, and hence the orbital separation
at which unipolar induction turns on.

Spin effects on the NS structure cannot be neglected if the NS spin
period $P$ is $\mathcal{O}({\rm ms})$. Furthermore, for comparable
mass BH--NSs near the tidal disruption radius, NS spin effects on the
orbit will be non-negligible when $P$ is similar to the BH--NS
encounter timescale~\citep{Tichy:2011gw}. For example, a BH--NS
eccentric encounter with mass ratio $q=M_{\rm BH}/M_{\rm NS}=4$ (as
studied here) near a periapse of $r_p=10M$ has an interaction
timescale of $t_{\rm int}\simeq(r_p^3/M)^{1/2}\sim1.0(M_{\rm
  NS}/1.4M_\odot)\rm ms$ ($M$ is the system's total mass, 
and we use geometric units with $G=c=1$ throughout).

NSs in field BH--NS binaries may not commonly have $P=\mathcal{O}({\rm ms})$ near
merger. However, there are two reasons to think that the opposite may
hold for dynamical capture BH--NS mergers occurring in GCs: the pulsar
spin period distribution in Galactic GCs peaks in the milliseconds, and
millisecond pulsars (MSPs) have longer inferred magnetic dipole
spin-down timescales.

Of the 144 currently known pulsars in Galactic GCs, $\sim83\%$ have
periods less than 10 ms, $\sim55\%$ less than 5 ms, and $\sim12\%$ have 
periods less than 2.5 ms~\footnote{\url{www.naic.edu/~pfreire/GCpsr.html}\label{Footnote1}}.
This set includes PSR-J1748-2446ad---the fastest-spinning pulsar
known, with $P=1.396$ ms~\citep{Hessels:2006ze}.  
The theoretical explanation for this skew toward short periods is that
GCs favor the formation of low-mass X-ray binaries
(LMXB)~\citep{Verbunt1987}, which are thought to 
spin up the NS to ms periods through mass and angular momentum
transfer~\citep{Alparetal1982,Radhakrishnan1982}.

Assuming that pulsar spin-down is predominantly due to magnetic dipole
emission, both the magnetic field strength ($B$) and spin-down timescale
($t_{\rm sd}$) can be computed from observations of $P$ and its time
derivative $\dot{P}$. For $B$ the relation is~\citep{Bhattacharya19911}
\begin{eqnarray}
B\sim 1.6\times10^{8}{\rm \ G}\left[\frac{P}{2.5{\rm \ ms}}\frac{\dot P}{10^{-20}}\right]^{1/2},\nonumber
\end{eqnarray}
which for known GC MSPs with $\dot{P}>0$ gives typical values 
$B\sim10^{8}\mbox{---}10^{9}$G. Observations of X-ray oscillations of
accretion-powered MSPs in LMXBs, and pulsar recycling theory imply
$B$ in the range $3\times10^7\mbox{---}3\times10^8$G~\citep{LambYu2005}.  For 
$t_{\rm sd}$ the expression is~\citep{ZhangMezaros2001}
\begin{eqnarray}
t_{\rm sd}&\sim& 4 {\rm \ Gyr} \frac{I}{10^{45}\rm{g}\ cm^2}\left(\frac{B}{3\times
10^8\rm{\ G}}\right)^{-2} \\ \nonumber
&&\left(\frac{P}{2.5 \rm \ ms}\right)^{2}\left(\frac{R_{\rm NS}}{10 \rm \
km}\right)^{-6},\nonumber
\end{eqnarray}
where the NS moment of inertia is $I$. 
Even neglecting the possibility of magnetic field decay, e.g. a pulsar
with a magnetic field of $3\times10^8$G and initial $P=2.5$ms 
will take roughly a Hubble time for $P$ to double.
Given the long spin-down timescale of MSPs and the results of~\citet{lee2010}, 
which suggest that in GCs there could be 40 BH--NS
collisions per Gyr per Milky Way-equivalent galaxy, it is at least
conceivable that some of these eccentric BH--NS collisions take place
with millisecond NSs. 

With this motivation, here we focus on eccentric BH--NS mergers (with initial
conditions corresponding to a marginally unbound
Newtonian orbit) and explore NS spin effects.
We show that even moderate spins
can significantly impact the outcome, both in terms of the GWs, 
and amounts of tidally stripped bound and unbound matter.  The
remainder of the paper is as follows: in
Sec.~\ref{numerical_approach} we describe our initial data and
numerical methods. In Sec.~\ref{results_and_discussion} we present 
our simulation results and discuss the impact of NS spin on
gravitational and EM signatures. We summarize in
Sec.~\ref{conclusions} and discuss future work.

\section{Numerical approach}
\label{numerical_approach}

We perform GR-HD simulations of BHs merging with rotating NSs using
the code of~\cite{code_paper}. The field equations are solved in the
generalized-harmonic formulation, using finite differences, while the
hydrodynamics are evolved using the same high-resolution
shock-capturing techniques as in~\cite{bhns_astro_paper}.

To construct initial data, we solve the constraint equations using the
code of~\cite{idsolve_paper}, specifying the free-data as a
superposition of a non-spinning BH with an equilibrium,
uniformly rotating NS, which we generate using the code
of~\cite{1994ApJ...424..823C,1994ApJ...422..227C}. For the NS EOS, we
adopt the HB piece-wise polytrope from~\cite{read}, and include a
thermal component $P_{\rm th}=0.5\epsilon_{\rm th}\rho$ allowing for
shock heating.

Fixing the NS gravitational mass to $1.35M_\odot$, we consider NSs
with dimensionless spins $a_{\rm NS}=J/M_{\rm NS}^2=0$,  0.1, 0.2, 0.3,
0.4, 0.5, and 0.756, having corresponding compactions (mass-to-equatorial-radius) 
$C=0.172$, 0.171,
0.169, 0.166, 0.161, 0.154, and $0.12$. The ranges of $T/|W|$ 
(kinetic-to-gravitational-potential-energy ratio) and $P$ in our spinning
NS models are $[0.003,0.12]$ and $[5.25,1.00]$ ms. The
fastest spinning NS considered has a polar-to-equatorial-radius ratio
of $r_{po}/r_{eq}=0.55$, near the mass-shedding limit of
$r_{po}/r_{eq}=0.543$.
We also vary 
$r_p/M\in[5,8]$.

We consider systems with $q=4$. The two COs are initially placed at a
separation of $d=50M$ ($\sim500$km), with positions and velocities
corresponding to a marginally unbound Newtonian orbit labeled by
$r_p$. For this initial study, we only consider cases where the spin
is aligned or anti-aligned with the orbital angular momentum (the
latter indicated by $a_{\rm NS}<0$).

\begin{figure} 
\begin{center} \hspace{-0.5cm}
\includegraphics[height=2.5in,clip=true,draft=false]{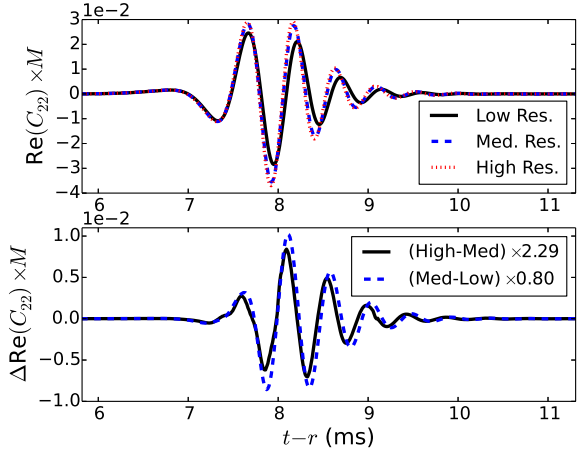}
\caption{ Convergence of the GW emission for the $r_p/M=6$, $a_{\rm NS}=0.756$ case.  
The top panel shows the real part of the $\ell=m=2$ mode of the Newman--Penrose scalar
$\Psi_4$ at three resolutions, and the bottom panel the differences in
this quantity with resolution, scaled assuming second-order
convergence.} \label{conv_plot}
\end{center} 
\end{figure}

The simulations utilize seven levels of adaptive mesh refinement
that are dynamically adjusted based on the estimated truncation error
of the metric.  Most simulations are performed using a base-level
resolution with $193^3$ points, and finest-level resolution with
approximately $75$ ($130$) points covering the (non-spinning) NS (BH)
diameter, respectively.  For $r_p/M=6$, $a_{\rm NS}=0.756$ we also
perform simulations at $2/3$ and $4/3\times$ the resolution, to
establish convergence and estimate truncation error. In
Fig.~\ref{conv_plot} we demonstrate the convergence in the GW emission.

\section{Results and discussion}
\label{results_and_discussion}

\subsection{Simple estimates}
\label{simple_estimates}

During eccentric encounters between NSs and BHs, for NS tidal
disruption to form a substantial disk and unbind a non-negligible
amount of matter, it must occur outside the ISO radius ($r_{\rm
  ISO}=4M_{\rm BH}$ for a marginally unbound test particle about a
non-spinning BH). 
In addition to shifting the effective ISO, spin makes the NS less self-bound, 
and thus alters the tidal disruption radius.
Equating the sum of the tidal and centrifugal accelerations to
the gravitational acceleration on the NS surface yields
\begin{equation}
\label{rtidal}
\frac{r_{t}}{M_{\rm BH}}\sim q^{-2/3}C^{-1}\left(\frac{1}{1-(a_{\rm NS}/a_{\rm ms})^2}\right)^{1/3},
\end{equation}
where we replaced the NS angular frequency with $\Omega=J/I=a_{\rm
  NS}M_{\rm NS}^2/I$, and let $I=2fM_{\rm NS}R_{\rm NS}^2/5$, with $f$
an order-unity constant that depends on the NS structure. Here,
$a_{\rm ms}=(4f^2/25 C)^{1/2}\simeq 0.8(f/0.7)(C/0.12)^{-1/2}$ is the
mass-shedding limit spin parameter. 
Equation~\eqref{rtidal} shows that the closer $a_{\rm NS}$ is
to $a_{\rm ms}$, the larger the tidal disruption radius. It also
suggests that for sufficiently fast rotators, the tidal disruption
radius can be outside the ISO even for large $q$, something which is
not true for non-spinning NSs unless the BH has near-extremal spin.
Additionally, as the prograde NS spin increases, the effective ISO
decreases. Therefore, we expect more massive disks and more unbound
material following tidal disruption outside the ISO with increasing
$a_{\rm NS}$.

\subsection{Dynamics and Gravitational Waves\label{Sec:GWs}} 
\begin{figure} 
\begin{center} 
\includegraphics[height=2.5in,clip=true,draft=false]{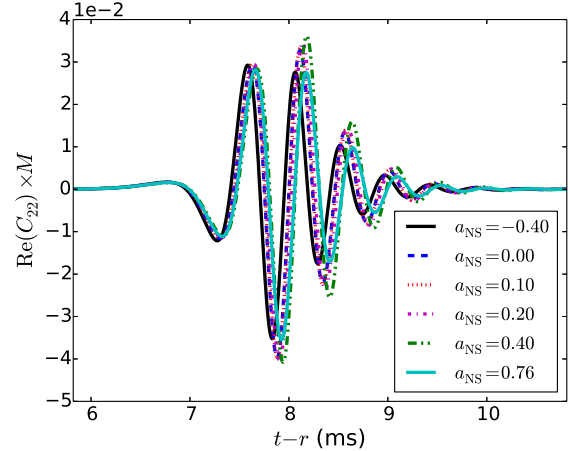}
\includegraphics[height=2.5in,clip=true,draft=false]{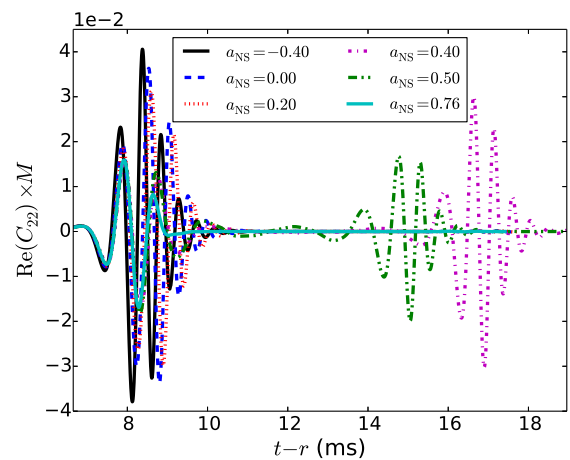}
\includegraphics[height=2.5in,clip=true,draft=false]{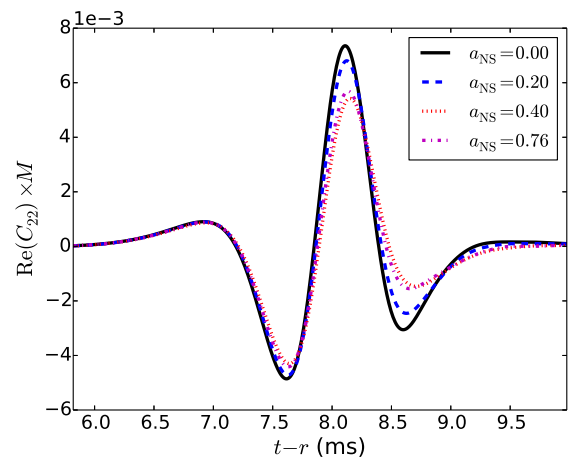}
\caption{Plots of the real part of the $l=m=2$ mode of the
  scalar $\Psi_4$. The top and middle panels show GWs
  from simulations with $r_p/M=6$ and 7, respectively. The
  bottom panel plots the GWs for $r_p/M=8$ fly-by cases.}
  \label{GWs_plot}
\end{center}
\end{figure}

\begin{figure*} 
\begin{center} 
\includegraphics[height = 1.275in]{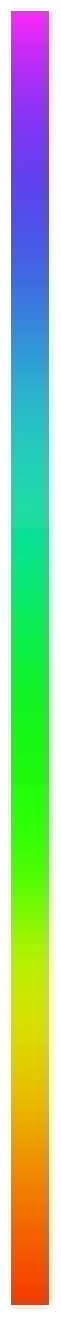}
\put(1,86){$10^{15}$ gm cm$^{-3}$}
\put(1,1){$10^{9}$}
\hspace{0.8 in}
\includegraphics[trim =5.5cm 4.61489cm 5.5cm 4.61489cm,height=1.275in,clip=true,draft=false]{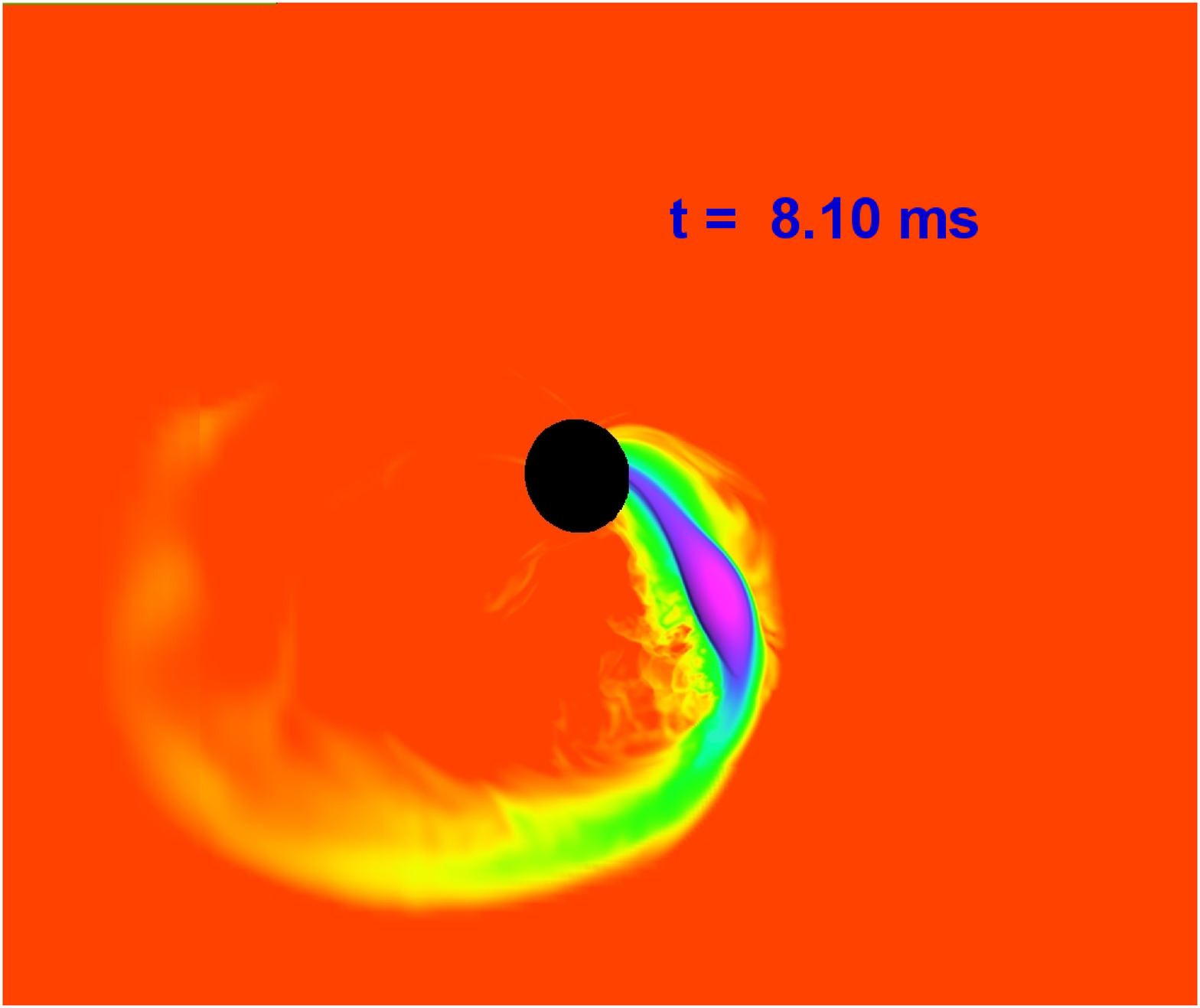}
\includegraphics[trim =5.5cm 4.61489cm 5.5cm 4.61489cm,height=1.275in,clip=true,draft=false]{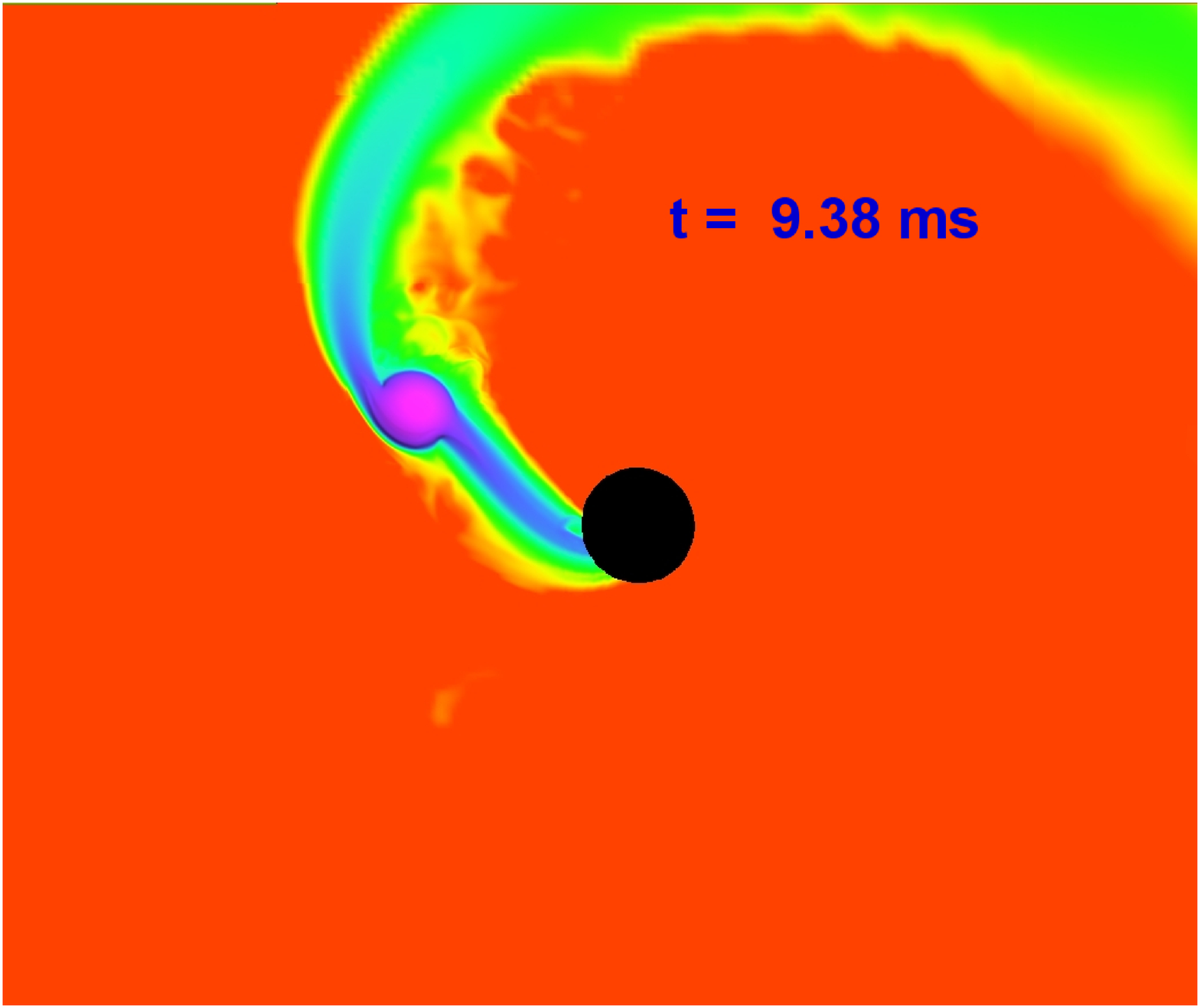}
\includegraphics[trim =5.5cm 4.61489cm 5.5cm 4.61489cm,height=1.275in,clip=true,draft=false]{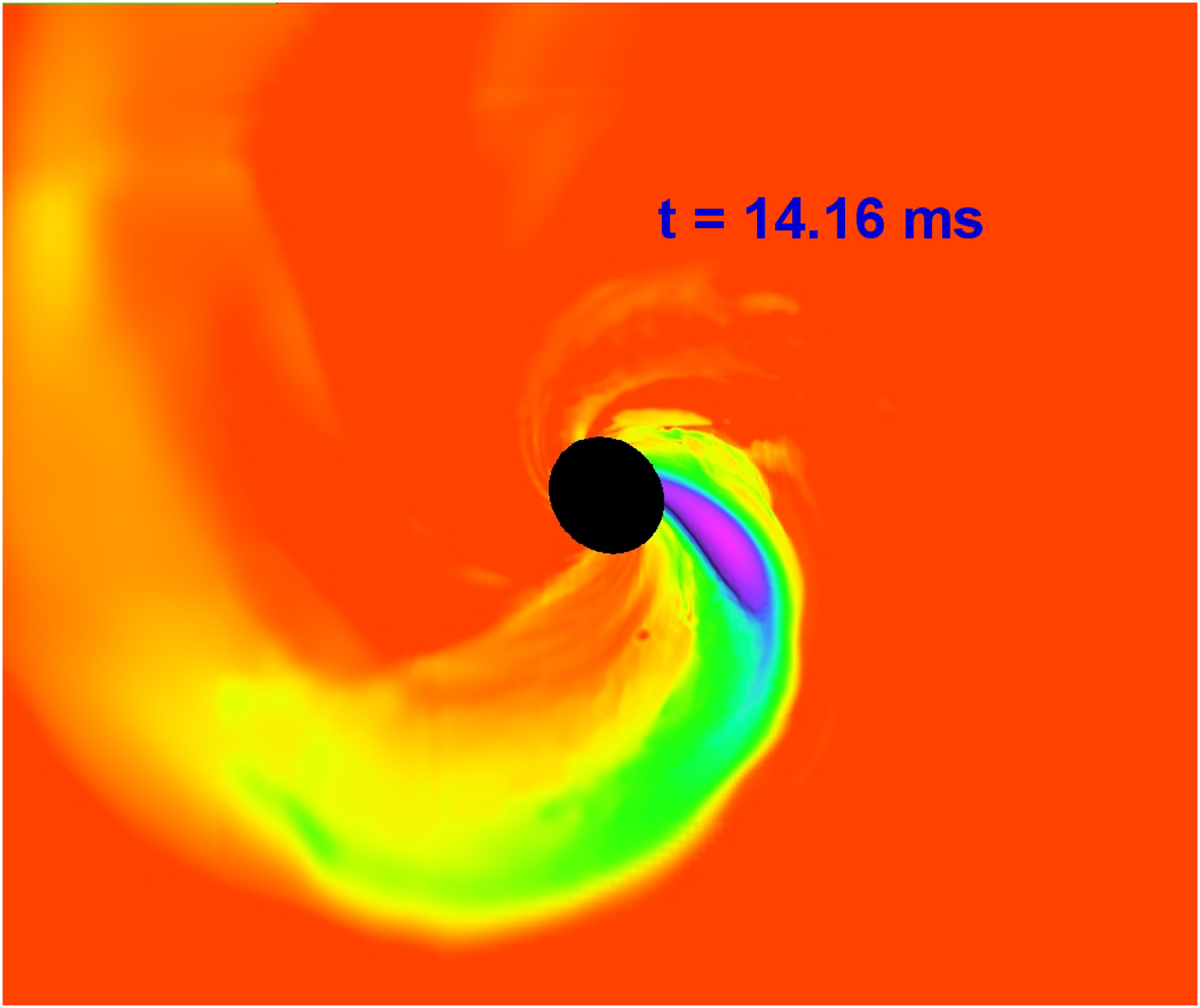}
\includegraphics[trim =5.5cm 4.61489cm 5.5cm 4.61489cm,height=1.275in,clip=true,draft=false]{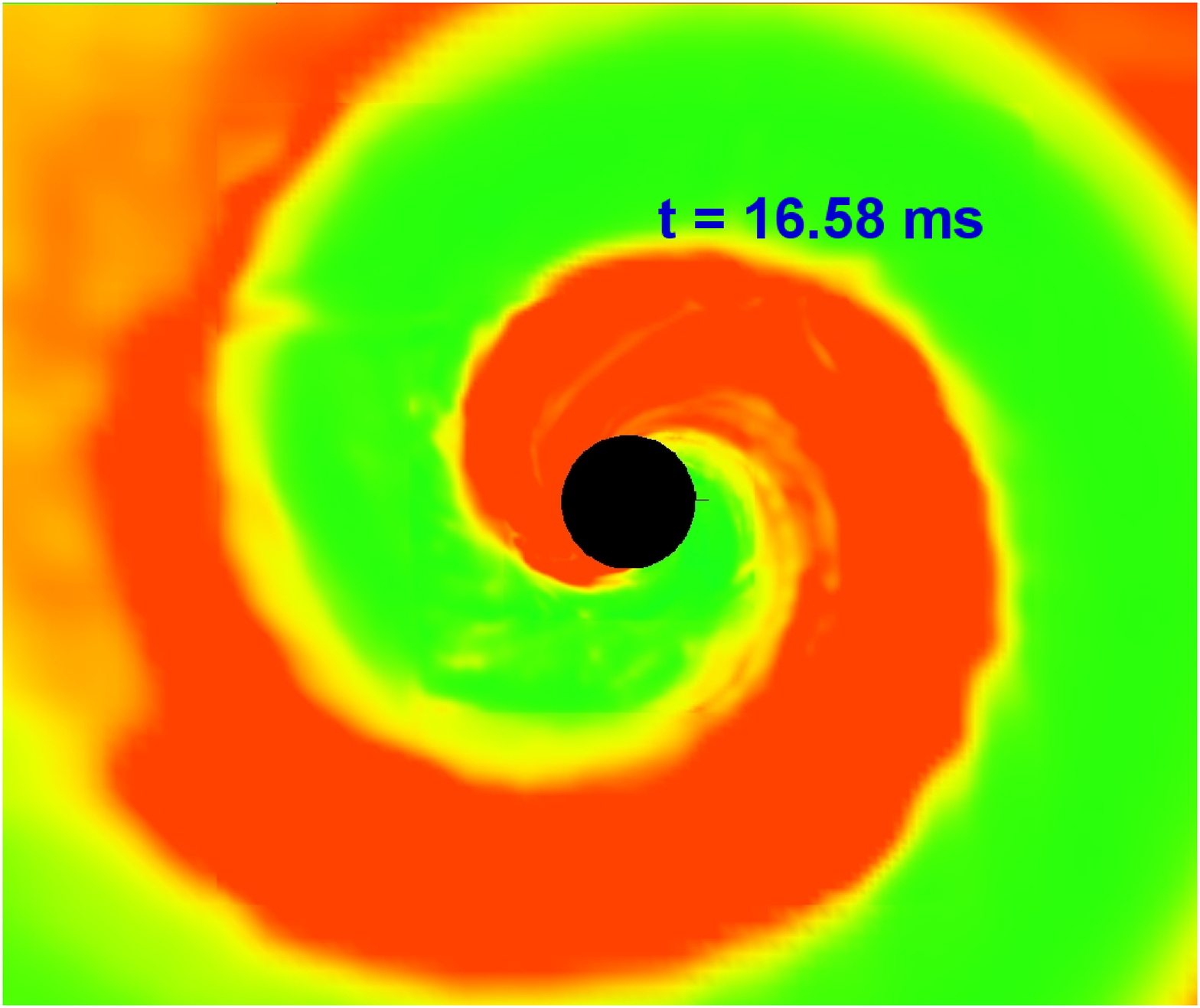}
\includegraphics[height = 1.275in]{vertical_scale.eps}
\put(1,86){$10^{15}$ gm cm$^{-3}$}
\put(1,1){$10^{9}$}
\hspace{0.8 in}
\includegraphics[trim =5.5cm 4.61489cm 5.5cm 4.61489cm,height=1.275in,clip=true,draft=false]{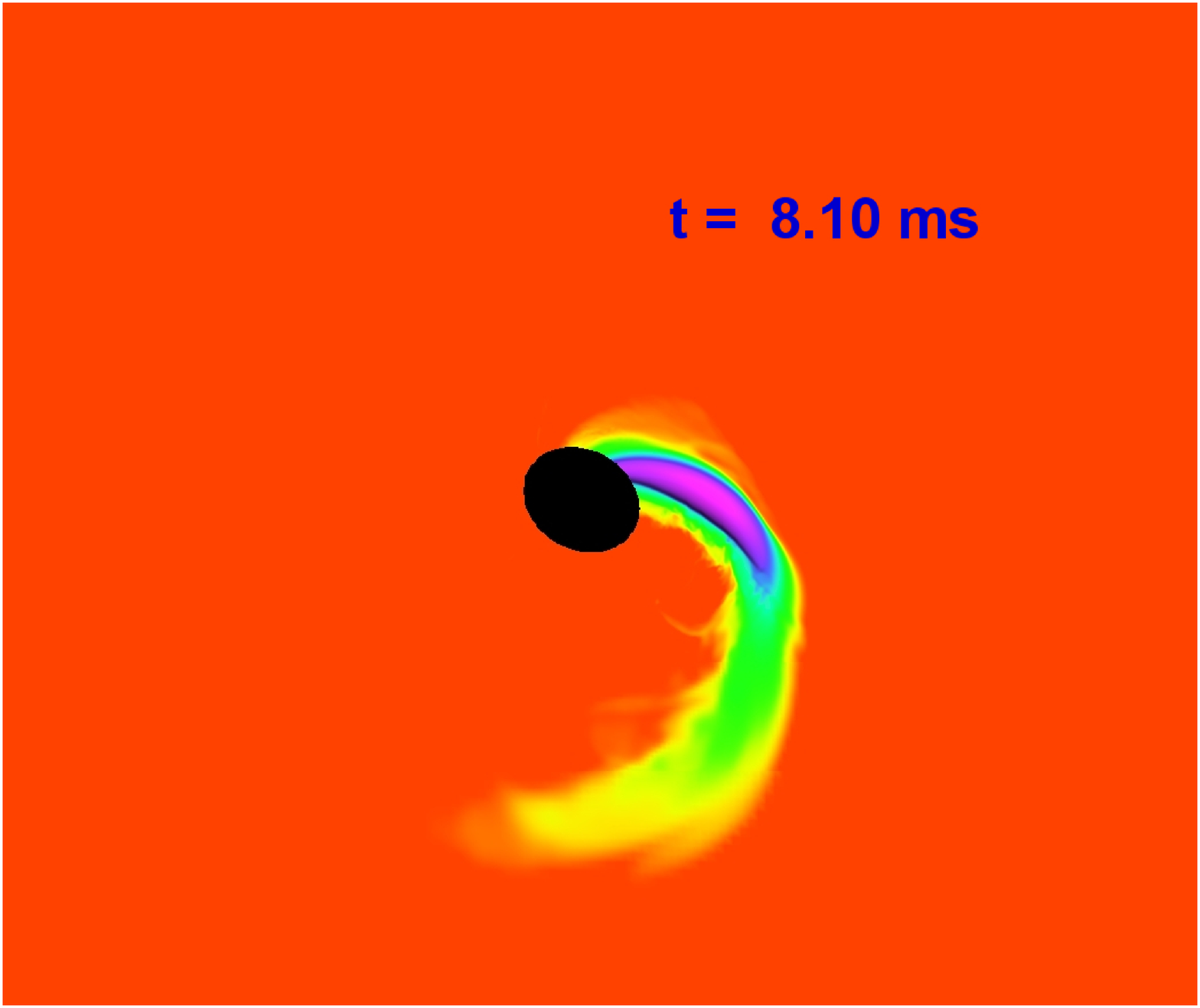}
\includegraphics[trim =5.5cm 4.61489cm 5.5cm 4.61489cm,height=1.275in,clip=true,draft=false]{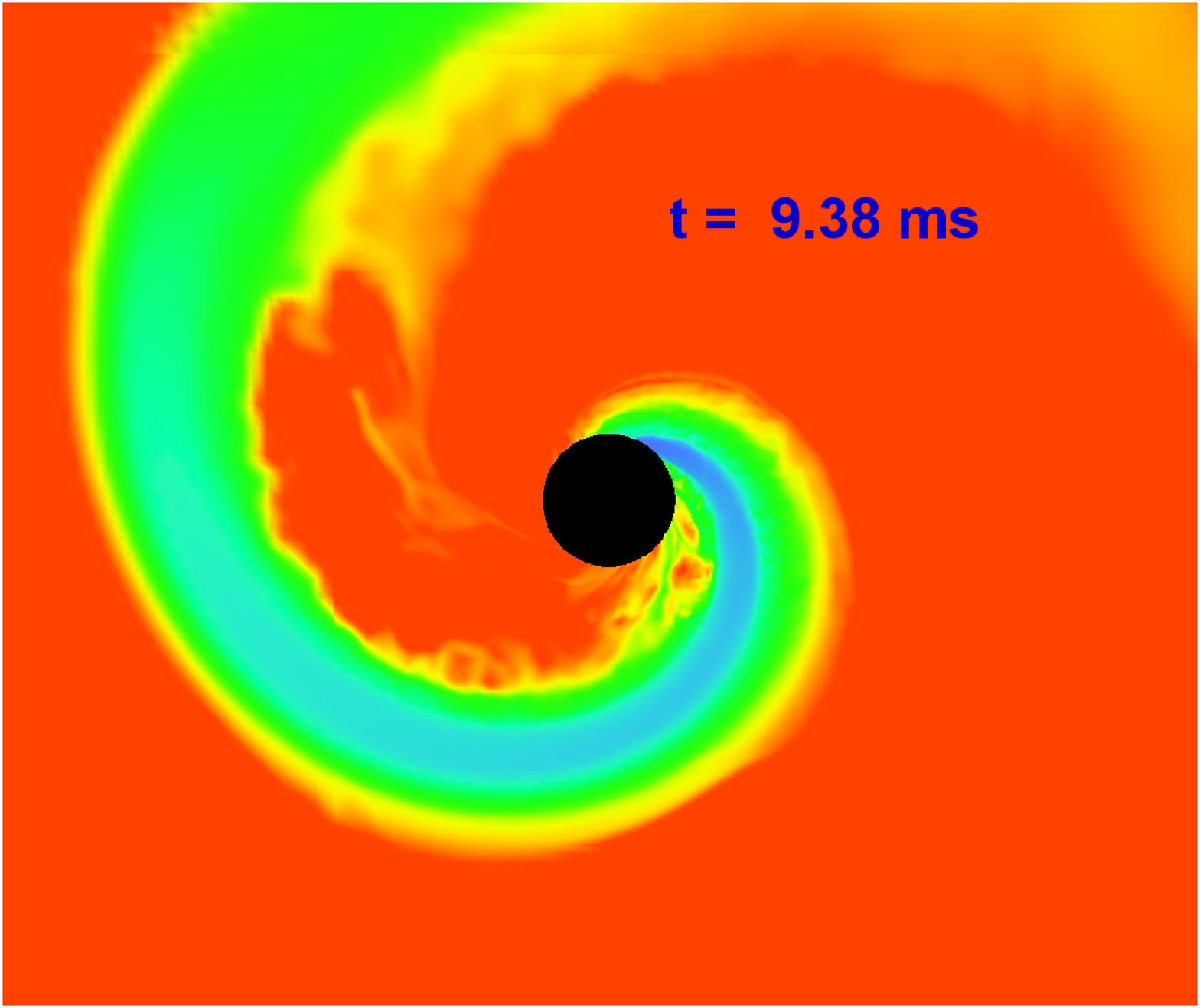}
\includegraphics[trim =5.5cm 4.61489cm 5.5cm 4.61489cm,height=1.275in,clip=true,draft=false]{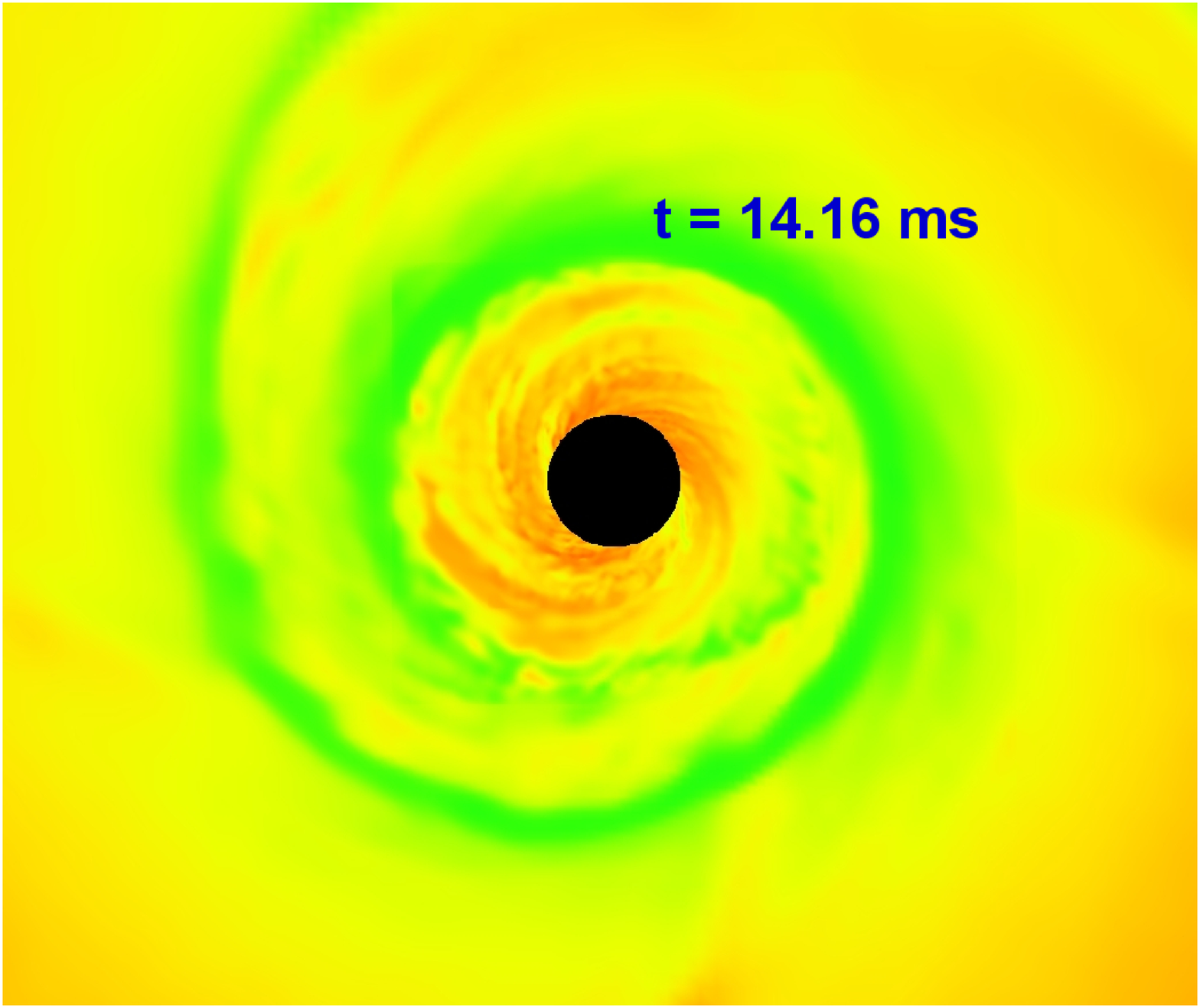}
\includegraphics[trim =5.5cm 4.61489cm 5.5cm 4.61489cm,height=1.275in,clip=true,draft=false]{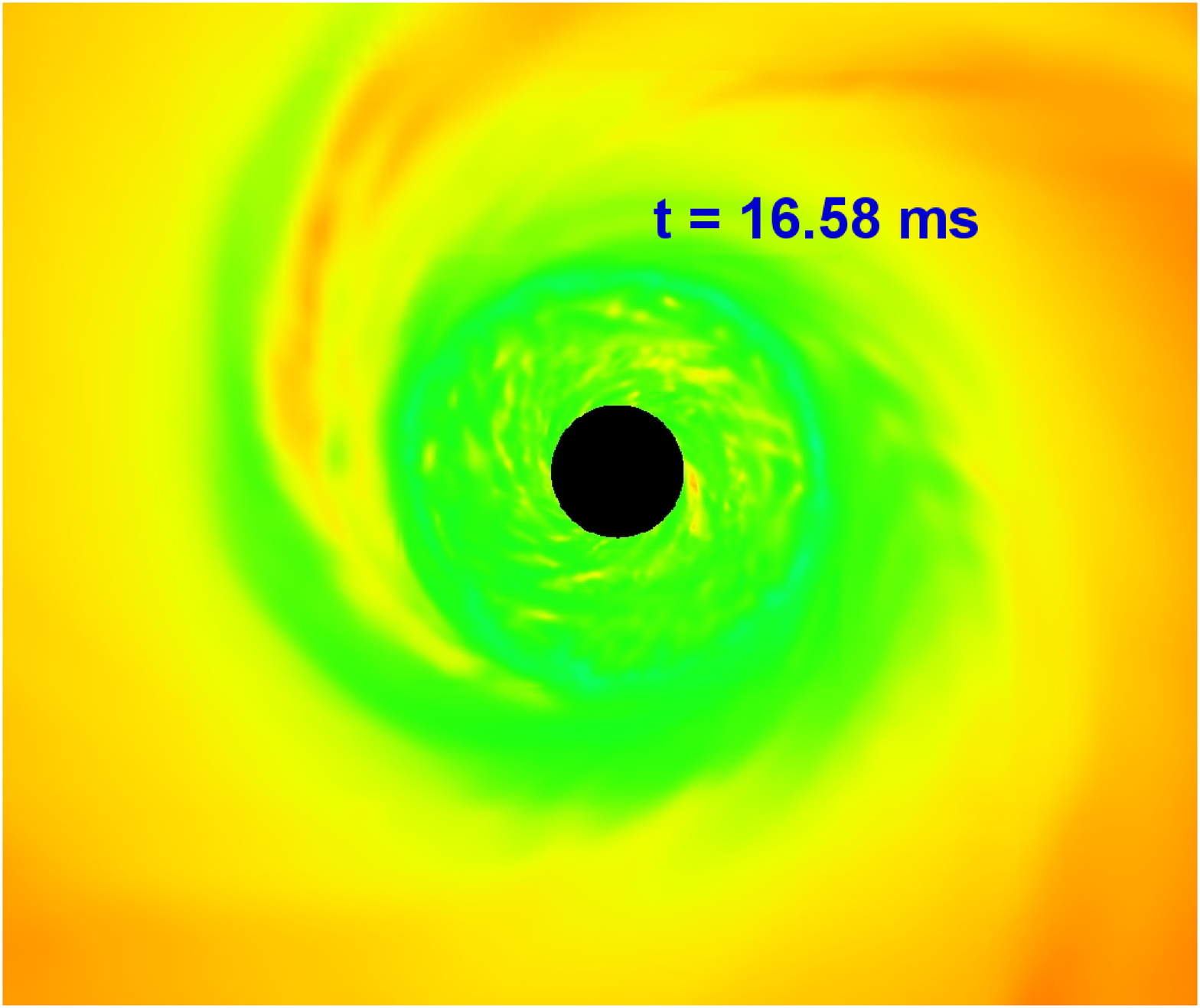}
\caption{Equatorial density snapshots. Top row ($r_p/M=7,\ a_{\rm
    NS}=0.5$) from left to right: the NS survives the first encounter
  (first and second panels), it is completely tidally disrupted during
  the second encounter (third panel), the bulk of the matter outside
  the BH is unbound (fourth panel).  Bottom row ($r_p/M=7,\ a_{\rm
    NS}=0.2$) from left to right: the NS is tidally disrupted during
  the first encounter (first panel), a tidal tail forms ejecting some
  matter to infinity (second panel), an accretion disk develops
  outside the BH (third and fourth panels). The scale can be inferred from the
  size of the BH ($R_{\rm BH}\sim16$ km).} \label{density_snapshots}
\end{center}
\end{figure*}

For the cases
considered here, those with $r_p\leq6.5$ merge on the initial
encounter, while those with $r_p\geq7.5$ go back out on an elliptic
orbit after fly-by.  Figure~\ref{GWs_plot} plots the dominant
contribution to the GW signal for $r_p/M=6$, 7, and 8; see also Table~\ref{bhrns_table}. 
Near the critical $r_p$ --- below (above) where a merger (fly-by)
occurs --- there are large differences in the dynamics that have a
noticeable impact on the GW signal and tidally stripped matter.  This
is evident here with the $r_p/M=7$ case, where there is either partial
tidal disruption followed by a merger on a second encounter (for
$a_{\rm NS}=0.4$ and 0.5), or complete tidal disruption/merger on the
first encounter (for the other spins).  This is illustrated in
Fig.~\ref{density_snapshots} for $a_{\rm NS}=0.2,0.5$, as well as the
bottom-right panel of Fig.~\ref{matter_unbound_plot} where it can be seen
that the NS for $a_{\rm NS}=0.4$ (0.5) loses $\sim10\%$ ($\sim20\%$)
of its mass in the initial encounter.

Though it is difficult to disentangle the nonlinear dynamics occurring
here, we suggest the following explanation for this non-monotonic
behavior as a function of $a_{\rm NS}$ for $r_p/M=7$.  The $a_{\rm
  NS}=0.756$ NS is the least bound and for this case complete tidal
disruption occurs on the initial encounter. As the NS is tidally
stretched, GW emission effectively shuts off (middle panel
Fig.~\ref{GWs_plot}), and matter begins to accrete onto the BH (bottom
panel of Fig.~\ref{matter_unbound_plot}). For the next two lower spins,
the material is more tightly bound, and only partial disruption
occurs, with some of the material promptly accreting onto the
BH. However, as the NS spin decreases, the effective ISO radius
increases, and for $a_{\rm NS}\lesssim0.3$ the core of the NS crosses
the ISO, resulting in immediate merger. The $a_{\rm NS}=0.4$ and 0.5
cases lose enough orbital energy and angular momentum during the first
encounter that they merge on the second.

Note that the simple estimate of Equation~(\ref{rtidal}) implies that
since at $r_p/M=7$ the $a_{\rm NS}=0.756$ case is completely
disrupted, while some of the lower spin cases are partially disrupted,
$r_p/M=8,a_{\rm NS}=0.756$ should also be disrupted, given how close
this spin is to the break up value. That this does {\em not} happen
shows that this crude estimate significantly underestimates the
self-binding of high spin stars.

\subsection{Post-merger Matter Distribution}

In Table~\ref{bhrns_table} we list the amount of bound and unbound mass exterior
to the BH shortly following 
merger. For $r_p/M=5,6$ the bound mass
is only a weak function of $a_{\rm NS}$, with the notable
exception of $a_{\rm NS}=0.756, r_p/M=6$. By contrast, near the
critical impact parameter ($r_p/M=7$) 
there is over an order of magnitude variation
in bound material as a function of NS spin.

The amount of bound rest-mass that forms a BH accretion disk is
$0.01\mbox{--}0.15M_\odot$ in our set. If these disks power sGRBs on
timescales of $\sim0.2$s, the accretion rates will be
$\sim0.05\mbox{--}0.75M_\odot\ \rm s^{-1}$, i.e., consistent with
magnetohydrodynamic BH--NS
studies~\citep{Paschalidis:2014qra}. Assuming a $1\%$ conversion
efficiency of accretion power to jet luminosity, these accretion rates
imply luminosities of $10^{51}\mbox{--}10^{52}\rm erg\ s^{-1}$---consistent with
characteristic sGRB luminosities.

\begin{figure*}
\begin{center}
\includegraphics[height=2.5in,clip=true,draft=false]{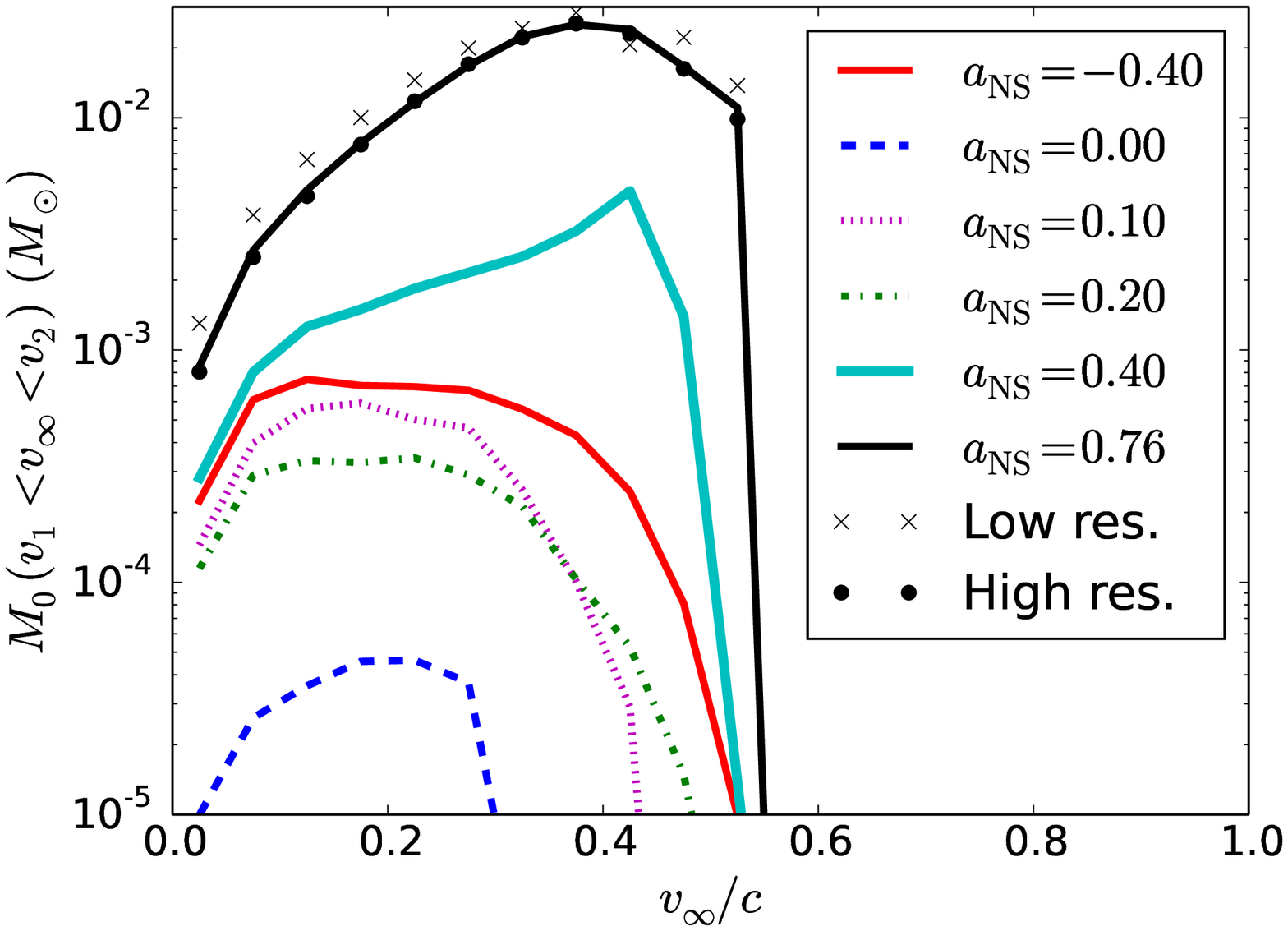}
\includegraphics[height=2.5in,clip=true,draft=false]{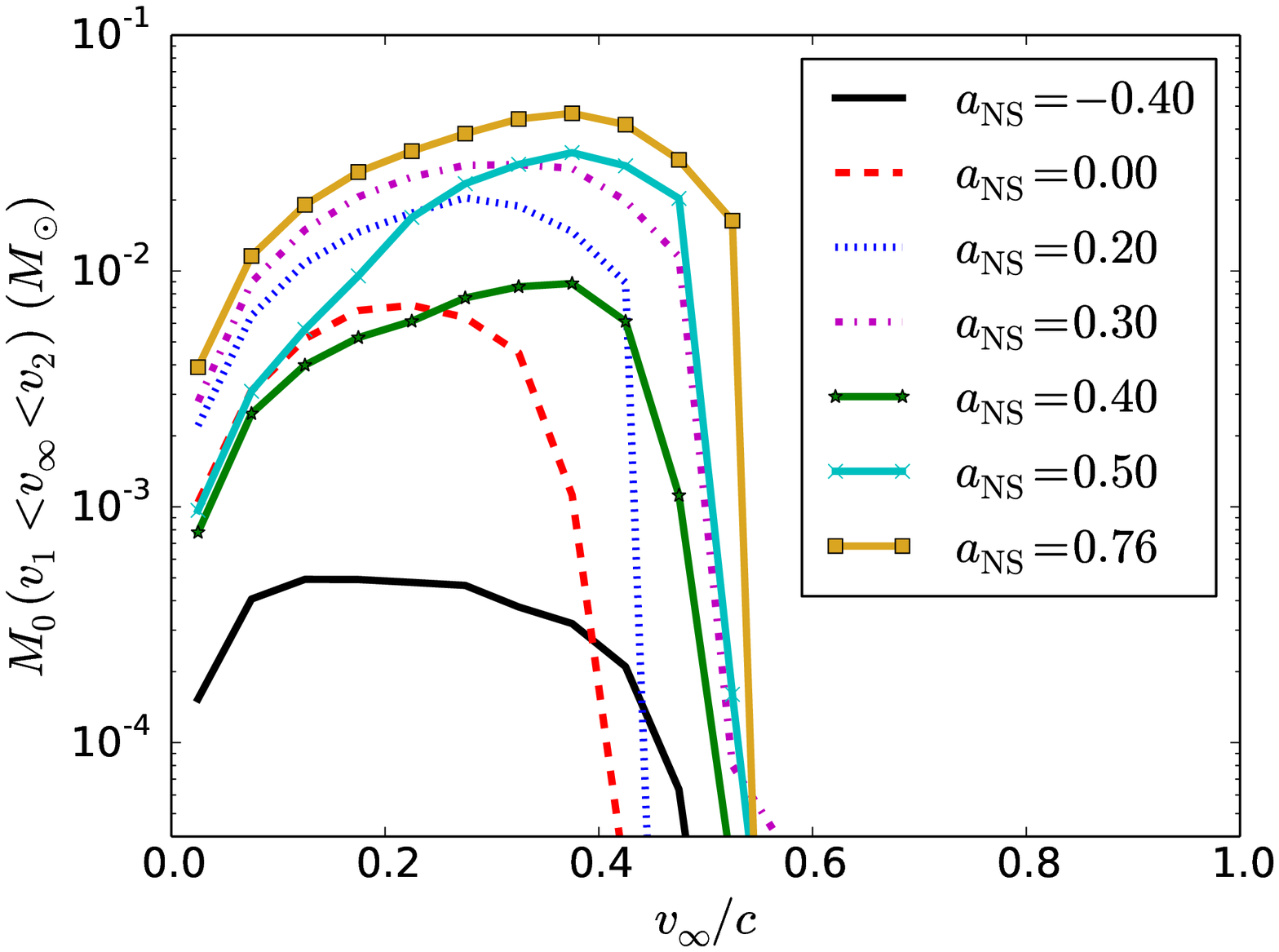}
\includegraphics[height=2.5in,clip=true,draft=false]{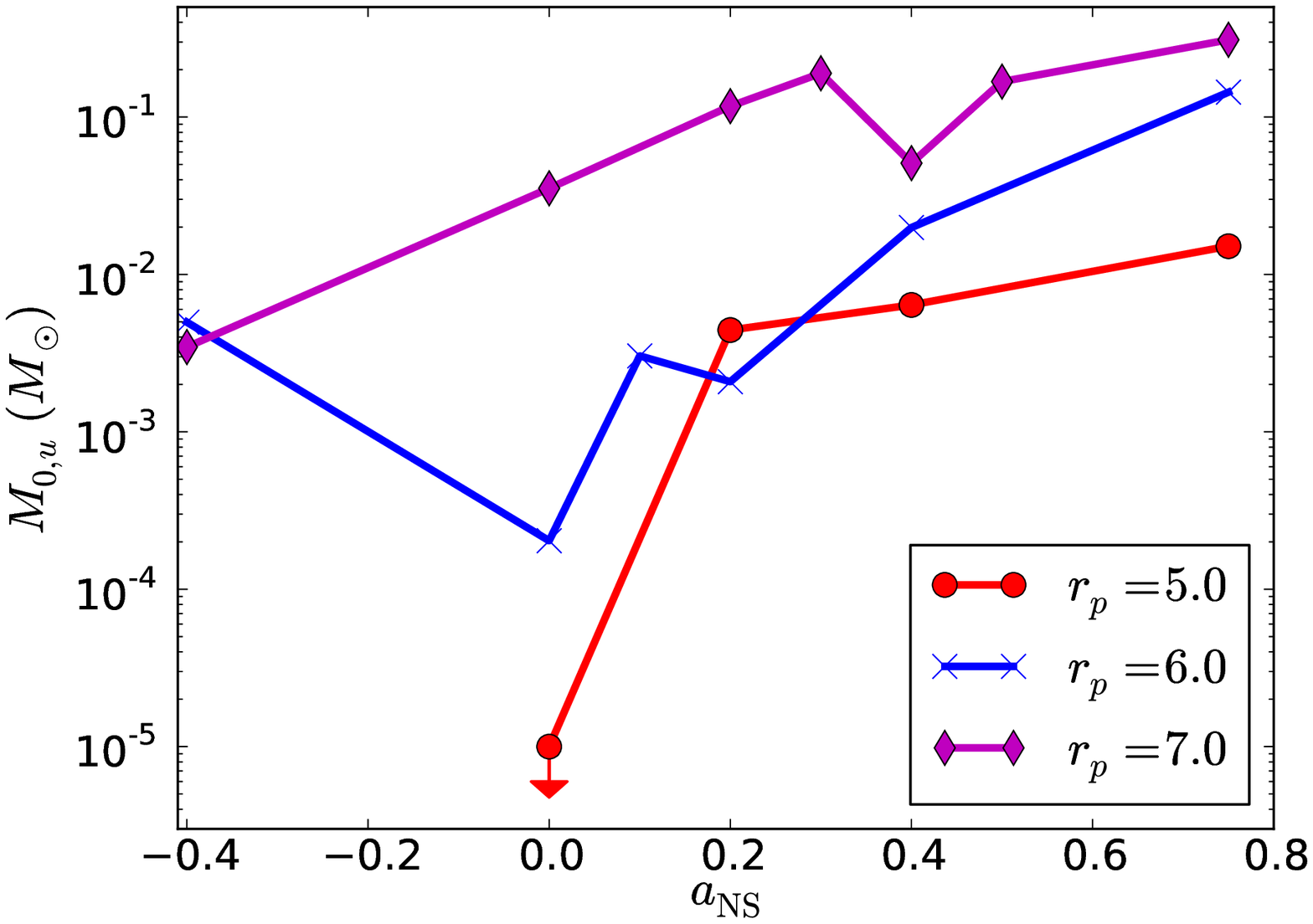}
\includegraphics[height=2.5in,clip=true,draft=false]{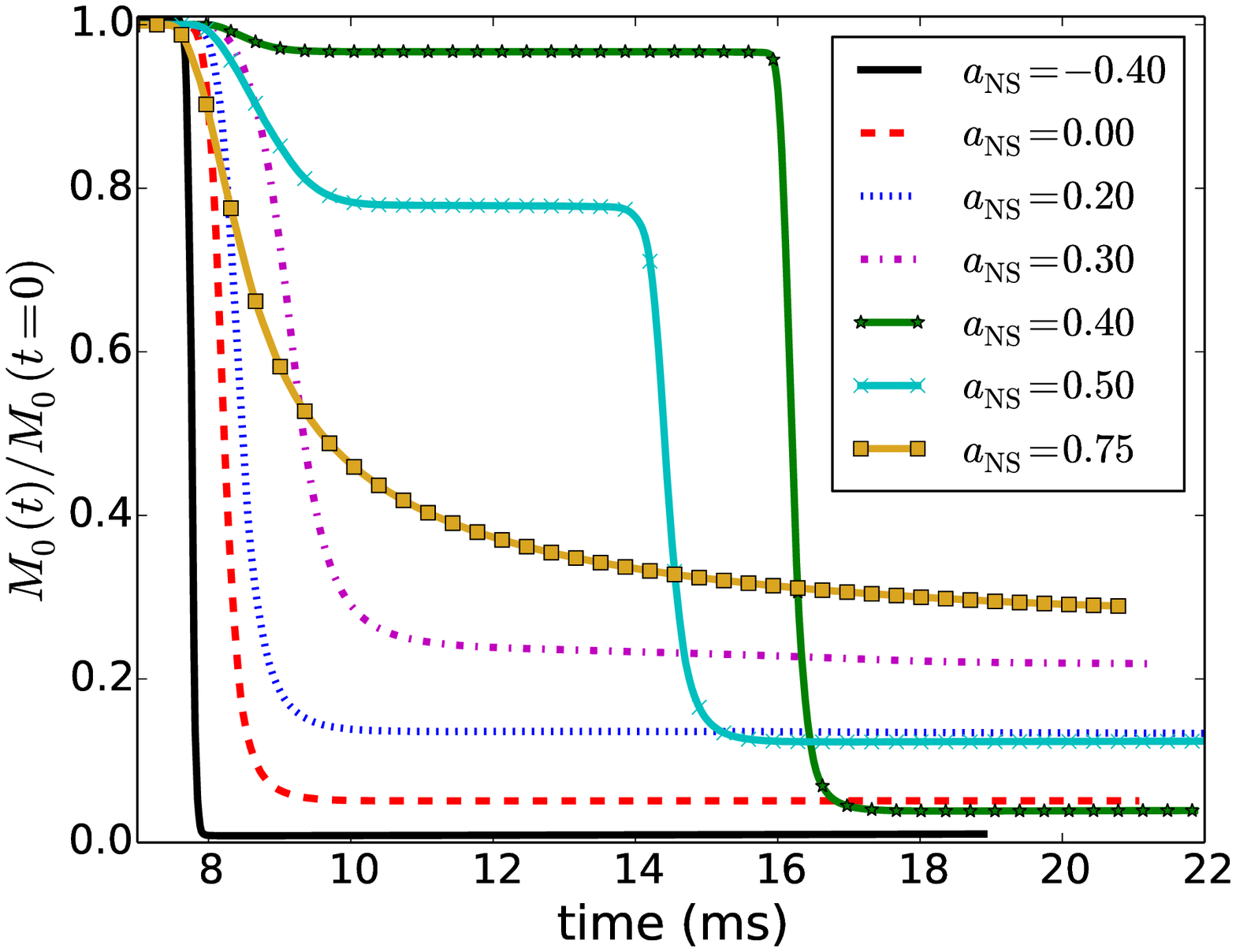}
\caption{ Top: distribution of the asymptotic velocity of unbound
  rest-mass, binned in increments of $0.05c$, and computed $\approx10$
  ms post-merger for $r_p/M=6$ (left) and $r_p/M=7$ (right) and
  various spins.  Bottom: total unbound mass as a function of NS spin (left)
  and rest-mass outside the BH versus time for $r_p/M=7$ and various spins
  (right).  Note that for the lowest point in the bottom-left panel, no unbound matter
  was found in the simulation (indicated by an arrow).
} \label{matter_unbound_plot}
\end{center}
\end{figure*}

The top two panels in Fig.~\ref{matter_unbound_plot} show plots of the
asymptotic velocity distribution of the unbound matter for
$r_p/M=6,7$. As anticipated (see Sec.~\eqref{simple_estimates}), the
general trend is that increasing $a_{\rm NS}$ increases both the
amount and average asymptotic velocity of unbound material. This is
also seen in Table~\ref{bhrns_table} where we list
these quantities for all cases. 
The bottom-left panel of Fig.~\ref{matter_unbound_plot} also demonstrates
that including spin increases the amount of unbound material by an
order of magnitude or more for the cases considered.

\begin{table*}
\begin{center}
{\scriptsize
\begin{tabular}{l l l l l l l l l l l l}
\hline 
$r_p$ &
$a_{\rm NS}$ &
$\frac{J_{\rm ADM}}{M^2}$\tablenotemark{a} &
$\frac{E_{\rm GW}}{M}\times100 $ \tablenotemark{b} &
$ \frac{J_{\rm GW}}{M^2}\times100$ \tablenotemark{c} & 
$M_{0,b}$ \tablenotemark{d} &
$M_{0,u}$ \tablenotemark{e} &
$<v_{\infty}>$ \tablenotemark{f} &
$E_{{\rm kin},51}$ \tablenotemark{g} &
$L_{41}$ \tablenotemark{h} &
$F_{\nu}$ \tablenotemark{i} &
$a_{\rm BH}$ \tablenotemark{j} 
\\
\hline
5.0   &   0.00 & 0.52 &   0.71   &   4.33   &  1.11   &   0.00   &   0.0  & 0.0   &   0.0   &   0.0  & 0.50 \\
5.0   &   0.20 & 0.52 &   0.73   &   4.51   &  0.96   &   0.44   &   0.20 & 0.09   &   1.1   &   0.02  & 0.51 \\
5.0   &   0.40 & 0.53 &   0.81   &   4.95   &  0.94   &   0.64   &   0.19 & 0.3   &   1.3   &   0.05  & 0.51 \\
5.0   &   0.75 & 0.55 &   0.81   &   4.97   &  1.12   &   1.51   &   0.42 & 3.0   &   2.9   &   4.5  & 0.51 \\
6.0   &  -0.40 & 0.55 &   0.92   &   5.66   &  1.01   &   0.50   &   0.22 & 0.3   &   1.2   &   0.08  & 0.51 \\ 
6.0   &   0.00 & 0.56 &   1.13   &   6.86   &  1.15   &   0.02   &   0.18 & 0.007 &   0.2   &   0.001  & 0.52 \\
6.0   &   0.10 & 0.57 &   1.16   &   7.47   &  1.07   &   0.30   &   0.19 & 0.1   &   0.9   &   0.02  & 0.53 \\
6.0   &   0.20 & 0.57 &   1.21   &   7.30   &  1.10   &   0.21   &   0.20 & 0.1   &   0.7   &   0.02  & 0.54\\
6.0   &   0.40 & 0.58 &   1.31   &   8.03   &  0.91   &   1.99   &   0.32 & 2.3   &   2.9   &   1.6  & 0.53\\
6.0 &   0.75 & 0.60 &   1.04(1.23)\tablenotemark{k}& 6.85(7.92)& 2.40(2.23)&14.39(14.01)   &   0.346(0.350) & 19.9   &   8.2   &   17.9     & 0.491 (0.492) \\
6.5   &   0.40 & 0.60 &   1.51   &   9.91   &  4.39   &   9.08   &   0.28 & 8.0   &   5.8   &   3.8  & 0.50 \\
7.0   &   -0.40& 0.63 &   1.51   &   9.38   &  1.17   &   0.35   &   0.23 & 0.2   &   1.0   &   0.06 & 0.53\\
7.0   &   0.00 & 0.61 &   1.72   &   11.65  &  4.31   &   3.53   &   0.21 & 1.7   &   3.1   &   0.4 & 0.51\\
7.0   &   0.20 & 0.62 &   1.68   &   12.13  &  8.95   &   11.73  &   0.26 & 8.6   &   6.3   &   3.4 &  0.47 \\
7.0   &   0.30 & 0.62 &   1.52   &   12.75  &  15.34  &   18.98  &   0.28 & 17.3  &   8.5   &   8.8 & 0.40  \\
7.0   &   0.40 & 0.63 &   2.12   &   18.27  &  0.94   &   5.10   &   0.28 & 4.6   &   4.4   &   2.3 & 0.45\\
7.0   &   0.50 & 0.63 &   1.65   &   15.39  &  2.06   &   16.80  &   0.33 & 20.4  &   8.6   &   16.1 & 0.50\\
7.0   &   0.75 & 0.64 &   0.70   &   6.95   &  12.43  &   30.98  &   0.32 & 36.6  &   11.4  &   25.3  & 0.37\\ 

\hline 

\end{tabular}
\caption{Summary of simulations followed through merger. \label{bhrns_table}} \begin{justify}
  For $r_p/M\geq7.5$ only the first fly-by encounter was modeled, hence
  no information related to disrupted material is available. The
  energy and angular momentum emitted in GWs for these cases drops
  with increasing $r_p$ after the first encounter, as expected. To
  within the estimated $20\%$ truncation error inferred from the
  $r_p=6$ case resolution study, we see no variation with
  spin. However, even a small variation in the energy
  emission at fly-by could result in a significant change in the time
  to the subsequent close encounter in a highly eccentric binary. Thus,
  higher resolution studies would be needed to ascertain the effect of
  spin on the GW signal for $r_p/M\geq7.5$.\end{justify}
\tablenotetext{1}{ADM angular momentum.}
\tablenotetext{2}{Total energy emitted in GWs through the $r=100 M$ surface.}
\tablenotetext{3}{Total angular momentum emitted in GWs.}
\tablenotetext{4}{Bound rest mass outside the BH $\sim 10$ ms post-merger in percent of $M_{\odot}$. }
\tablenotetext{5}{Unbound rest mass in percent of $M_{\odot}$. }
\tablenotetext{6}{Rest-mass averaged asymptotic velocity of unbound material.}
\tablenotetext{7}{Kinetic energy of ejecta in units of $10^{51}$ erg.}
\tablenotetext{8}{Kilonovae bolometric luminosity in units of $10^{41}\rm erg\ s^{-1}$ using Eq.~\eqref{Lkilonovae}.}
\tablenotetext{9}{Specific brightness from ejecta interaction with ISM in units of mJy using Eq.~\eqref{Fnu}. 
}
\tablenotetext{10}{Remnant BH dimensionless spin.}
\tablenotetext{11}{Values in parentheses are Richardson extrapolated values
using all three resolutions.}
}
\end{center}

\end{table*}

The increase in total rest-mass $M_{\rm 0,u}$ and velocity $v$ of the unbound
material with increasing $a_{\rm NS}$ can strongly impact potential kilonovae
signatures from such mergers. These arise when neutron-rich ejecta produce
heavy elements through the r-process than then undergo fission, emitting
photons~\citep{Li:1998bw,2005astro.ph.10256K}.  
Recently~\cite{2013ApJ...775...18B} have shown that the
opacities in r-process ejecta will likely be dominated by
lanthanides, giving rise times of 
\[t_{\rm peak}\approx0.25 (M_{\rm
0,u}/10^{-2}\ M_{\odot})^{1/2}(v/0.3c)^{-1/2} \ \mbox{ d}\nonumber\]
with peak luminosities of
\begin{equation}L\approx 2\times 10^{41}\left(\frac{M_{\rm
0,u}}{10^{-2}\ M_{\odot}}\right)^{1/2}\left(\frac{v}{0.3c}\right)^{1/2}\mbox{ erg
s$^{-1}$}\label{Lkilonovae}\end{equation} 
for typical values found here.  In some cases, opacities an order of
magnitude lower than those used above may be
justified~\citep{2015MNRAS.446.1115M}.  Using Eq.~\eqref{Lkilonovae}
we estimate the luminosity from potential kilonovae in
Table~\ref{bhrns_table}.  Apart from the fact that NS spin can make
the difference in whether there will be a kilonova at all for
$r_p/M=5$, for $r_p/M=6$ and $r_p/M=7$ spin affects $L$ by an order of
magnitude in our set. For $r_p/M=6$ even a moderate $a_{\rm NS}=0.1$
increases $L$ by a factor of 4 compared to $a_{\rm
  NS}=0$. \citet{2013ApJ...775...18B} predict that a kilonova
luminosity of $\sim10^{41} \rm erg\ s^{-1}$ corresponds to an r-band
magnitude of 23.5 mag at 200 Mpc (near the edge of the aLIGO volume),
above the planned LSST survey sensitivity of 24.5 mag.
Thus, differences in luminosity by factors of a few could be discernible.

Ejecta will also sweep the interstellar medium (ISM) producing radio
waves. These will peak on timescales of weeks with
brightness~\citep{2011Natur.478...82N}
\begin{eqnarray} 
  F(\nu_{\rm obs}) &\approx& 0.6(E_{\rm kin}/10^{51}\mbox{
    erg})(n_0/0.1{\rm \ cm}^{-3})^{7/8} \label{Fnu} \\ 
     && (v/0.3c)^{11/4}(\nu_{\rm obs}/{\rm GHz})^{-3/4}(d/100{\rm \ Mpc})^{-2}\mbox{ mJy}\nonumber 
\end{eqnarray} 
for an observation frequency $\nu_{\rm obs}$ at a distance $d$, and
using $n_0\sim0.1$ cm$^{-3}$ as the density for GC
cores~\citep{2013MNRAS.430.2585R}.  Estimating the kinetic energy and
the mass-averaged velocity in the ejecta, we show $F(\nu_{\rm
  obs})$ via Eq.~\eqref{Fnu} in Table~\ref{bhrns_table}. For $r_p/M=6$
($r_p/M=7$) $F(\nu_{\rm obs})$ varies by 3 (2)
orders of magnitude over our set of spins.

Finally, it has been suggested that ejecta from mergers involving NSs
may make a non-negligible contribution to the overall abundance of
r-process elements ~\citep{1974ApJ...192L.145L,Rosswog:1998gc}.
In particular,
dynamical-capture binaries, which can form and merge on shorter
timescales, may be favored over field binaries in explaining
abundances in carbon-enhanced metal-poor
stars~\citep{2014arXiv1410.3467R}. The average galactic production of
these elements is estimated to be $\sim10^{-6}M_{\odot}$
yr$^{-1}$~\citep{2000ApJ...534L..67Q}.  Making the limiting assumption
that all r-process material comes from extreme BH-NS mergers cases
like the $r_p=7$, $a_{\rm NS}=0.76$ with $M_{0,\rm
  u}\approx0.3M_{\odot}$ caps these extreme events at $3\times10^{-6}$ yr$^{-1}$
per galaxy (similar to predicted rates for primordial BH--NS
mergers~\citep{2010CQGra..27q3001A}). 
 
\section{Conclusions}
\label{conclusions}

We have demonstrated using GR-HD simulations of dynamical capture
BH--NS mergers that even moderate values of NS spin can significantly
increase the mean velocity and amount of unbound material (to as much as
$0.3M_{\odot}$ for extreme spins).  This could lead to
significantly brighter transients, including kilonovae a factor of a
few brighter, and radio wave emission from interaction with the ISM an
order of magnitude or more brighter.  For comparison, simulations of
quasicircular BH--NS mergers with nonspinning NSs typically find ejecta
velocities $\sim0.2\mbox{--}0.3c$, comparable, though somewhat smaller
than found here, but only find similar amounts of ejected material for
cases with smaller mass-ratios and/or high BH
spin~\citep{2015arXiv150205402K}.  We also find that the NS spin can
alter the amount of bound matter that, following tidal disruption,
remains to form an accretion disk that may power a sGRB. Depending on
the impact parameter and NS spin, these mergers can produce accretion
disks of up to a tenth of a solar mass.

We find that near the critical impact parameter the NS spin influences
the orbital dynamics to a sufficient extent to affect whether a merger
or fly-by occurs, with a corresponding large effect on the GW
emission.  At a first glance this variability might seem exceedingly
rare, requiring a finely tuned impact parameter. However, since the
primary source of this sensitivity to binary parameters arises because
the pericenter gets close to the region of unstable orbits, which
exists for all eccentricities, not merely the initially hyperbolic 
case considered here, one can speculate that the last few encounters
for any case where non-negligible orbital eccentricity remains will be
subject to this sensitivity.  Likewise, the variability associated
with EM counterparts could also be present for a larger range of
initial impact parameters. Future simulations of multi-burst events
will be needed to address this speculation.  At the other end of the
spectrum, some fraction of dynamical-capture binaries that form at
larger initial separations will circularize prior to merger due to GW
emission; the results found here thus also motivate the study of
quasicircular mergers involving millisecond NSs.

We have shown it is important to include spin to understand the full
range of possible EM and GW outcomes in eccentric mergers. 
However, whether it will be possible to perform parameter estimation from a 
putative multimessenger event is a different question. Certainly in a single 
burst event the
degeneracies will be too strong to, for example, 
identify NS spin as the sole reason for an unusually bright
counterpart. Multi-burst events can in principle lift much of the
degeneracy, as information in the timing of the bursts could
significantly narrow the parameters of the progenitor binary. The
range of viable NS EOSs, NS spin directions, and BH spins needs
to be simulated, both to determine how these parameters affect the
observable outcomes, and how they add to or lift degeneracies. GW detection
rates and parameter estimation also needs to be investigated within a
realistic data analysis framework including detector noise. All of
these problems we leave for future studies.  We also plan to study the
effect of spin in dynamical-capture NS--NS mergers.

\acknowledgments

We are grateful to Stuart Shapiro for access to the equilibrium
rotating NS code. This work was supported by NSF grant PHY-1305682 and
the Simons Foundation. Computational resources were provided by
XSEDE/TACC under grant TG-PHY100053 and the Orbital cluster at
Princeton University.


\bibliographystyle{hapj}
\bibliography{ref}

\end{document}